\documentclass[10pt, twocolumn,english,amssymb,aps,superscriptaddress,showpacs,amsmath,showkeys,floatfix,pra]{revtex4-2}
\usepackage[T1]{fontenc}
\usepackage[latin9]{inputenc}
\usepackage{geometry}
\usepackage[active]{srcltx}
\usepackage{amsmath}
\usepackage{graphicx}
\usepackage{esint}
\usepackage{epsfig}
\usepackage{epstopdf}
\usepackage{physics}
\usepackage{color}
\usepackage{natbib}
\usepackage{upgreek}
\usepackage{mathtools}
\usepackage{soul}
\usepackage{lipsum}
\usepackage{enumitem}
\usepackage{amsfonts}
\usepackage{eurosym}
\usepackage{braket}
\usepackage{setspace}
\usepackage{siunitx}
\usepackage[normalem]{ulem}

\newcommand\identity{1\kern-0.25em\text{l}}

\newcommand{\atcm}{\mathrm{at.cm}^{-3}}
\usepackage[dvipsnames]{xcolor}
\usepackage{bm}
\usepackage{ulem}
\usepackage{color}
\definecolor{blue}{rgb}{0,0,1}
\definecolor{green}{rgb}{0,1,0}
\definecolor{purple}{rgb}{0.5,0,1}
\makeatletter
\usepackage{babel}
\makeatother

\begin{document}

\title{Non-linear density scaling of spin noise reveals atomic correlations in warm vapors}

\author{J. Delpy}
\affiliation{Universit\'e Paris-Saclay, CNRS, \'Ecole Normale Sup\'erieure Paris-Saclay, CentraleSup\'elec, LuMIn, Orsay, France}
\affiliation{Department of Applied Physics, University of Geneva, 1205 Geneva, Switzerland}
\author{E. Cardoz}
\affiliation{Universit\'e Paris-Saclay, CNRS, \'Ecole Normale Sup\'erieure Paris-Saclay, CentraleSup\'elec, LuMIn, Orsay, France}

\author{K.V. Adwaith}
\affiliation{Universit\'e Paris-Saclay, CNRS, \'Ecole Normale Sup\'erieure Paris-Saclay, CentraleSup\'elec, LuMIn, Orsay, France}

\author{N. Fayard}
\affiliation{Universit\'e Paris-Saclay, CNRS, \'Ecole Normale Sup\'erieure Paris-Saclay, CentraleSup\'elec, LuMIn, Orsay, France}

\author{N. Belabas}
\affiliation{Centre de Nanosciences et de Nanotechnologies, CNRS, Universit\'e Paris-Saclay, 91120 Palaiseau, France}

\author{F. Bretenaker}
\affiliation{Universit\'e Paris-Saclay, CNRS, \'Ecole Normale Sup\'erieure Paris-Saclay, CentraleSup\'elec, LuMIn, Orsay, France}

\author{F. Goldfarb}
\affiliation{Universit\'e Paris-Saclay, CNRS, \'Ecole Normale Sup\'erieure Paris-Saclay, CentraleSup\'elec, LuMIn, Orsay, France}

\begin{abstract}

We experimentally demonstrate a non-linear dependence of the spin noise variance on atomic density in a warm alkali vapor. Implementing high-bandwidth spin noise spectroscopy (SNS) near the $\mathrm{D}_2$ transition of rubidium, a quadratic spin noise contribution is shown to arise at high densities,
in contrast with the linear dependence valid in non-interacting ensembles.
This non-linear scaling is shown to crucially depend on the residual optical excitation of the vapor by the probe beam, suggesting it stems from atomic cross-correlations due to resonant dipole-dipole interaction (DDI) in the vapor. We support this claim by introducing an additional experimental protocol to quench the ddi, resulting in a suppression of both the quadratic scaling of the spin variance and the distortions of the spin noise spectrum induced by the interaction.
These results extend the applications of SNS to the characterization of many-body correlations in complex quantum systems.

\end{abstract}

\maketitle
\section{Introduction}

Spin properties lie at the heart of numerous quantum technologies, from metrology sensors to electronic devices. Some of them, such as optically-pumped magnetometers (OPM) \cite{bloom_principles_1962, budker_optical_2007}, atomic clocks \cite{schmittberger2020reviewcontemporaryatomicfrequency,bauch2003caesium} or quantum storage platforms \cite{Novikova2007, tittel_photon-echo_2010,Sangouard}, rely on the manipulation of non-interacting spin ensembles. In contrast, the second quantum revolution is led by technologies exploiting non-classical correlations between particles \cite{PRXQuantum.1.020101}, such as quantum communication and cryptography \cite{scarani2009security, xu2020secure}.

A powerful experimental method to probe spin dynamics in a wide variety of systems is the so-called spin noise spectroscopy (SNS) \cite{aleksandrov_magnetic_1981}. Optically measuring spontaneous fluctuations of the total spin in an ensemble allows to characterize its structural and coherence properties. SNS was successfully performed in non-interacting systems such as atomic vapors \cite{crooker_spectroscopy_2004, mihaila_quantitative_2006, katsoprinakis_measurement_2007, liu_birefringence_2022}, bulk and low-dimensional semiconductors \cite{oestreich_spin_2005, hubner_rise_2014, poltavtsev_spin_2014, glasenapp_spin_2016}, and solid-state spins \cite{kozlov_spin_2023, kozlov_spin_2025}. This method reveals both the fundamental mechanisms and the limitations in such systems by identifying noise and decoherence channels, even far from equilibrium \cite{glasenapp_spin_2014, swar_measurements_2018, sun_non-equilibrium_2022, delpy_spin-noise_2023,sato_fluctuations_2025}. This proves useful when engineering for instance quantum sensors such as OPM, in which spin noise often limits sensitivity \cite{shah_high_2010, mouloudakis_spin_2024,li_spin_2025}, promoting the use of atomic and light squeezed states \cite{lucivero_squeezed-light_2016,troullinou_squeezed-light_2021,guarrera_spin-noise_2021}.

Nevertheless, implementations of SNS in correlated systems are scarce. Spin fluctuations detected by optical interferometry revealed entanglement in an ultracold Fermi gas a decade ago \cite{meineke_interferometric_2012}, yet the current SNS protocol could only detect artificially driven spin noise in cold atoms \cite{swar_detection_2021}. In alkali vapors, a frequency redistribution across spin noise spectra was shown to originate from spin-exchange collisions \cite{mouloudakis_effects_2022, mouloudakis_interspecies_2023, roy_cross-correlation_2015}, although no signatures of atomic correlations were found in the total spin noise variance. The technical challenge of implementing SNS in more complex systems persists to this day, despite several theoretical proposals suggesting that SNS could be used to probe spin interactions \cite{gorshkov_phase_2024}, magnetic phase transitions \cite{schlegel_spin-noise_2024} or many-body localization \cite{roy_probing_2015}. 

To contribute bridging the gap between non-interacting and strongly-correlated ensemble, we study SNS in a warm rubidium vapor. We reported in Ref.\,\cite{delpy_anomalous_2025} distortions of spin noise spectra originating from dipole-dipole interactions, arising between atoms in the ground state and a residual excited-state population induced by the probe light. These previous observations clearly indicated the breakdown of the single-atom description, but the narrow detection bandwidth prevented an unambiguous observation of many-body correlations. In the present work, we aim at providing a clear proof of such correlations by measuring the density scaling of the absolute spin noise variance, whose deviation from linearity directly reflects correlated spin fluctuations. 

This paper is organized as follows: section \ref{section1} introduces the experimental setup used to perform SNS measurements with  a high bandwidth over two orders of magnitude of the vapor density. We report a quadratic scaling of the spin noise variance with density occurring at high temperatures. In section \ref{section2}, we analyze the dependence of this anomalous scaling on the residual optical vapor excitation, by investigating the impact of the probe laser parameters. This leads us to interpret our observations in term of correlated atomic spin noise induced by optical interactions. Finally, section \ref{section3} supports this interpretation by introducing a novel optical method to quench dipole-dipole interactions and measure subsequent changes in the spin noise variance.

\section{Experimental measurement of non-linear spin noise variance}\label{section1}
\subsection{Experimental setup and method}\label{section1_setup}
\begin{figure}
    \centering
    \includegraphics[width=\columnwidth]{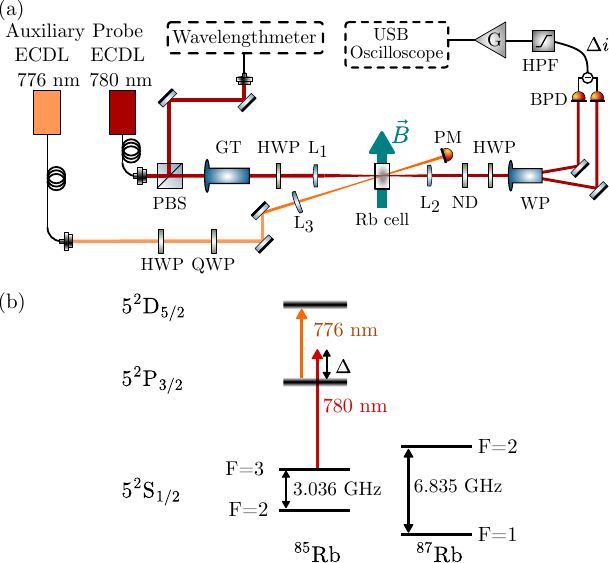}
    \caption{\textbf{Experimental setup and energy structure of rubidium.} (a) Schematics of the experimental setup. The 780 nm probe beam is linearly polarized using a Glan-Thompson (GT) polarizer before entering the 1-mm rubidium vapor cell. Output polarization fluctuations are measured using a balanced detection scheme comprising a half-wave plate (HWP), a Wollaston prism (WP) and a balanced photodiode (BPD). The detected power is adjusted using a variable neutral density filter (ND). A magnetic field $\vec{B}$ is applied transversely. The role of the auxiliary 776 nm laser (depicted in orange) is discussed in section \ref{section3}. $L_1,\, L_2,\, L_3$: focusing and collimation lenses, QWP: quarter-wave plate, PBS: polarizing beamsplitter, PM: powermeter. A bias-tee acts as a high-pass filter (HPF), G denotes a low-noise amplifier. (b) Diagram of energy levels involved in this study. $F$ numbers denote ground-state hyperfine levels. The laser detuning $\Delta$ is defined in the main text.}
    \label{figure1}
\end{figure}

A schematic of the experimental setup is shown in Fig.\,\ref{figure1}(a).
We measure Faraday rotation (FR) fluctuations due to spin noise (SN) in a rubidium vapor by tuning an external cavity diode laser (ECDL, Toptica DL pro) near the D$_2$ line of rubidium atoms. A Glan-Thompson polarizer produces a high-purity linear polarization of the probe light, which is then focused to a diameter of $\displaystyle{\phi \simeq 300~\unit{\micro\meter}}$ inside the vapor cell. The latter consists of a 1-mm-thick glass cell filled with natural rubidium, without any buffer gas or anti-relaxation coating. The cell is embedded in a metallic oven and heated using two 15 W resistors.
Two T-type thermocouples measure the temperature in the vicinity of both the probed area and the Rb reservoir, with a $\pm2^{\circ}\mathrm{C}$ estimated precision. The first temperature $T$ is used to infer the vapor density using the Killian formula \cite{steck_rubidium_2001}. A home-made temperature controller acts as a switch to a DC power supply feeding current to the heating resistors. A magnetic field of 19~G corresponding to a Larmor frequency of $\simeq$10~MHz for $^{85}$Rb and $\simeq$15 MHz for $^{87}$Rb is applied in a Voigt configuration. This field shifts the spin-noise resonance from zero frequency to the corresponding Larmor frequencies of the two isotopes, above any dominating low-frequency technical noise \cite{crooker_spectroscopy_2004}. Other noises, including the detection chain background and laser shot noise, are appropriately subtracted as elaborated in appendix \ref{app_background}. The cell along with the oven is shielded from any ambient magnetic field using a three-layer $\mu$-metal shield.

This experimental setup allows to address densities spanning over two orders of magnitude. Even at the far detunings considered here, which are typically more than five times larger than the Doppler broadening of 250 to $320\,\mathrm{MHz}$ half width at half maximum (HWHM) in our experimental conditions, the absorption of the probe light becomes non-negligible at high density. To eliminate the dependence of the measured signal strength on the optical power incident on the photodetector, which varies with density because of absorption, we use a variable neutral density filter placed after the cell to keep the optical power reaching the detection stage constant. In the following, we refer to it as the "detected power", denoted $P_d$, to distinguish it from the input probe power $P_p$ in the cell. A Wollaston prism then splits the beam of power $P_d$ into its orthogonal polarization components that are balanced using a half-wave plate. The balanced beams are sent to a 350 MHz bandwidth balanced photodetector (Thorlabs PDB435A) that produces a signal equal to the difference between the photocurrents corresponding to the two beams, $\Delta i \propto i_+ - i_-$. This differential current is, in turn, proportional to the instantaneous fluctuations in the rotation angle of the probe, $\Delta i \propto \delta \theta \left( t \right)$ and therefore to the instantaneous spin fluctuations \cite{aleksandrov_magnetic_1981}. A bias-tee filters out frequency components below 200 kHz, while the AC component is amplified with a 23.8\,$\mathrm{dB}$ gain. This signal is fed into a USB oscilloscope (Pico Technology PicoScope 3418E MSO), with a 500 MHz analog bandwidth and a 1 GHz sampling rate, to digitize the data and compute the spin noise power spectral density (PSD). Finally, the spin noise variance is computed by integrating under the PSD \cite{katsoprinakis_measurement_2007}. The integration procedure is discussed in appendix \ref{app_background}.

\begin{figure}
    \centering
    \includegraphics[width=\columnwidth]{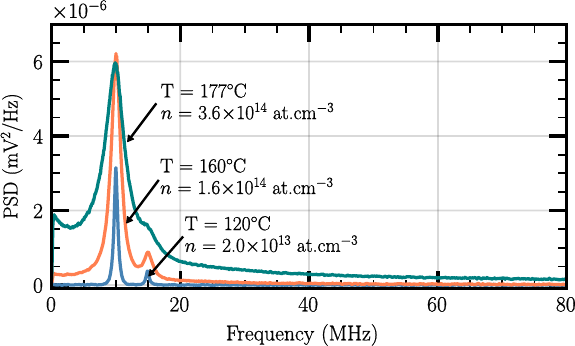}
    \caption{\textbf{Experimental spin noise spectra.} Examples of spin noise PSD obtained for densities below and above $10^{14}\,\atcm$, for a detuning $\Delta/2\pi = +1.5\,\mathrm{GHz}$ and a $2\,\mathrm{mW}$ input probe power.}
    \label{figure2}
\end{figure}

Figure \ref{figure2} shows typical spin-noise spectra at different atomic number densities obtained for a detuning $\displaystyle{\Delta/2\pi = +1.5\,\mathrm{GHz}}$ and an input probe power $P_p$ of 2.0~mW. Here, the detuning is defined as $\displaystyle{\Delta = \omega_p-\omega_0}$, with $\omega_p$  the probe laser frequency and $\omega_0$ the $\displaystyle{5^2\mathrm{S}_{1/2}, \, F=3 \rightarrow 5^2\mathrm{P}_{3/2}}$ transition frequency of $^{85}$Rb. The signal is detected at a fixed power $P_d=400\,\unit{\micro\watt}$ for all densities. The noise spectrum corresponding to $n\simeq2\times10^{14}\,\atcm$ displays a standard spin-noise resonance peaking at the respective Larmor frequencies of $^{85}$Rb and $^{87}$Rb. The main noise and broadening source here is the finite transit time of atoms through the probe beam. However, for densities exceeding $10^{14}\,\atcm$, a clear additional broadening of the resonance peaks is observed, accompanied by a broadband noise component that lifts the baseline dramatically, especially at lower frequencies. This is prominently seen at $\displaystyle{n\simeq3.6\times10^{14}\,\atcm}$. This long-tail component contributes to the total noise power up to frequencies as high as 200 MHz. These high density effects have previously been attributed to resonant dipole-dipole interaction in the vapor when the average inter-atomic separation is less the probe light's wavelength, i.e., $\langle r \rangle\lesssim\lambda / 2\pi$ \cite{delpy_anomalous_2025}. In the following, we study the effects of this interaction on the total spin noise variance, which carries information on the interplay between the stochastic evolution of each atom \cite{mouloudakis_interspecies_2023}.

\subsection{Non-linear density dependence of the absolute spin noise variance}\label{section1_exp}

\begin{figure}
    \centering
    \includegraphics[width=\columnwidth]{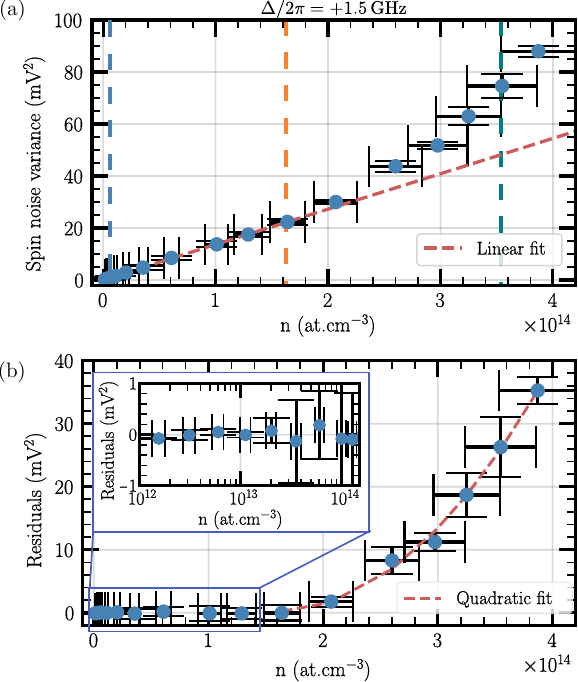}
    \caption{\textbf{Evolution of the absolute spin noise variance with vapor density.} (a) Evolution of the integrated spin noise variance as a function of atom density for a detuning $\Delta/2\pi = +1.5\,\mathrm{GHz}$, a $2\,\mathrm{mW}$ probe power and $0.4\,\mathrm{mW}$ detected power. Red dashed line: linear fit to the data up to $n=10^{14}\,\atcm$. Color dashed lines indicate the corresponding spin noise PSD in Fig.\,\ref{figure2}. (b) Residuals from the linear regression, fitted with a quadratic function for $n>1.5\times 10^{14}\, \atcm$ (red dashed line). Inset: close-up view of the low-density data points.}
    \label{figure3}
\end{figure}

Figure \ref{figure3} (a) shows results of total spin noise variance integrated from power spectra similar to the ones in Fig.\,\ref{figure2}, as a function of the vapor density, ranging from $\displaystyle{n=1.5\times 10^{12}\,\atcm}$ to $\displaystyle{3.9\times 10^{14}\,\atcm}$ ($\displaystyle{T=80^{\circ}\mathrm{C}}$ to $T=179^{\circ}\mathrm{C}$). The detuning is $\displaystyle{\Delta/2\pi = +1.5\,\mathrm{GHz}}$, corresponding to a probe laser tuned between the $\displaystyle{F=3\rightarrow F'}$ and $\displaystyle{F=2\rightarrow F'}$ transitions of Rubidium 85. The input probe power is $P_p=2\,\mathrm{mW}$, while the detected power is fixed at $P_d=400\,\unit{\micro\watt}$ for all densities. The data show two distinct regimes in the evolution of the spin noise, below and above a characteristic density $\displaystyle{n_c\simeq1.5\times 10^{14}\,\atcm}$. For $n<n_c$, the spin noise variance follows a linear trend up to densities of the order of $10^{14}\,\atcm$, as expected for non-interacting vapors \cite{crooker_spectroscopy_2004, katsoprinakis_measurement_2007}. In this regime, a simple linear regression (in red dashed line), yields a slope of $\kappa_{exp} = 0.138\pm0.02\,\mathrm{nV}^2.\mathrm{cm}^{3}$. Error bars on the spin noise variance are due to uncertainties in the background subtraction (see appendix \ref{app_background}), while errors on the density arise from uncertainties on the temperature inside the cell. To validate our spin noise and density measurements, we compare this value with a light-atom coupling model \cite{shah_high_2010, vasilakis_precision_2011}, accounting for the hyperfine structure of rubidium and our experimental parameters. Assuming uncorrelated atoms, we obtain a theoretical slope $\kappa_{th} = 0.130\,\mathrm{nV}^2.\mathrm{cm}^{3}$, in good agreement with the experimental slope $\kappa_{exp}$. The complete calculations can be found in appendix \ref{app_SN_calculations}. This confirms that the spin noise is consistent with that of independent rubidium atoms close to thermal equilibrium.

Strikingly, a strong deviation from this linear regime is then observed for densities $n>n_c$. The SN power visibly increases faster than expected, reaching absolute values almost twice as high at $\displaystyle{n= 3.9\times 10^{14}\,\atcm}$ than predicted by the previous linear scaling. This superlinear increase is further evidenced in Fig.\,\ref{figure3}(b), which shows the residuals from the linear fit at low densities. The fact that this deviation appears at a threshold rather than continuously is demonstrated by the inset figure, which shows that all data points agree with the linear regression within measurement uncertainties below $n_c$. The residuals at higher densities can then well be fitted by a quadratic function with a density offset $\displaystyle{f(n,n_c)=\gamma_{exp} (n-n_c)^2}$ (red dashed line), which is incompatible with the independent atom picture.

We note that a quadratic scaling of the spin projection variance has been observed in cold atomic clouds and Bose-Einstein condensates \cite{koschorreck_quantum_2010, koschorreck_sub-projection-noise_2010, shibata_dispersive_2025}, but were attributed to the coupling of light to residual spin polarization via tensorial light shifts. In our experiment, due to the unresolved excited-states hyperfine structure, the tensorial polarizability of the vapor is negligible for large detunings \cite{happer_effective_1967, happer_light_1970}. Moreover, the vapor being unpolarized, we can discard non-linearities due to probe back-action in our work. Finally, the quadratic spin noise cannot be imputed to the detection, since the detected power $P_d$ is weak enough for the photodetection to operate in a linear regime, as demonstrated in appendix \ref{app_linearity}.

Finally, the non-linear spin noise contribution seems to arise jointly with the distortions of the spin noise spectra discussed in section \ref{section1_setup} and Ref.\,\cite{delpy_anomalous_2025}. Such changes were attributed to resonant dipole-dipole interactions induced by the off-resonant probe beam. We therefore suggest that the superlinear spin noise variance evidenced here could have the same origin. Specifically, it could arise from positive correlations between optically interacting atoms at high density. In the next section, we support this claim by investigating the dependence of the non-linear spin variance on the residual excitation of the vapor by the probe beam.

\section{Characterization of optically-driven atom-atom spin noise correlations}\label{section2}

Resonant dipole-dipole interactions describe the coherent energy exchange between atoms lying in a ground state and the small excited state population induced by the detuned probe laser. This fraction is governed by the saturation parameter $\displaystyle{s\propto\dfrac{\Omega^2}{\Delta^2 + \Gamma_0^2/4}}$ \cite{Grynberg_Aspect_Fabre_2010}, where $1/\Gamma_0$ is the excited state lifetime and $\Omega$ the probe Rabi frequency. Therefore, we expect a dependence of the observed non-linear spin noise on the two parameters $\Omega$ and $\Delta$, which we now experimentally investigate.
\begin{figure}
    \centering
    \includegraphics[width=\columnwidth]{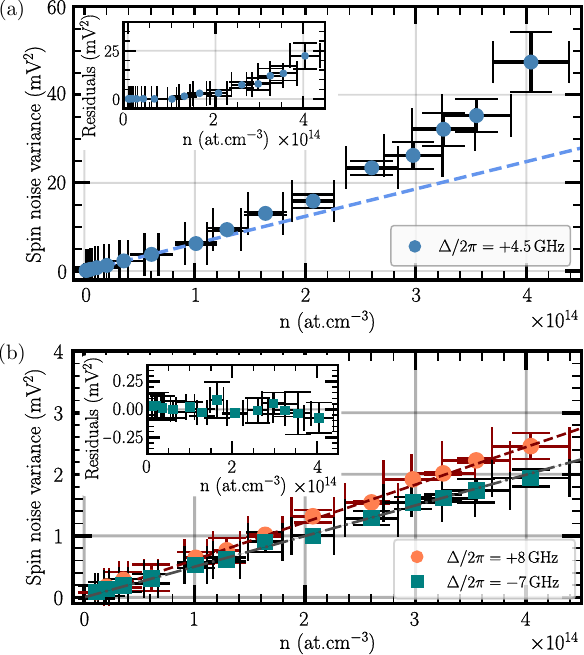}
    \caption{\textbf{Dependence of the spin noise scaling on the probe detuning.} Evolution of the spin noise variance as a function of vapor density for a detuning of (a) $\displaystyle{\Delta/2\pi=+4.5\,\mathrm{GHz}}$ and (b) $\displaystyle{\Delta/2\pi=-7\,\mathrm{GHz}}$ (green squares),  $\displaystyle{\Delta/2\pi=8\,\mathrm{GHz}}$ (orange dots). Dashed and dash-dotted lines are linear fit to the data for $n<10^{14}\,\atcm$. Insets: residuals from linear fits from respective color-coded data.}
    \label{figure4}
\end{figure}

\subsection{Impact of the optical detuning}
We first evidence the impact of the probe light on atom-atom interactions by performing the previous measurement at different probe laser frequencies. Figure \ref{figure4}\,(a) shows the evolution of the spin noise variance for a detuning $\displaystyle{\Delta/2\pi = +4.5\,\mathrm{GHz}}$, i.e. a probe laser tuned between the $\displaystyle{F=2\rightarrow F'}$ transitions of Rubidium 85 and $\displaystyle{F=1\rightarrow F'}$ of rubidium 87 (see Fig.\,\ref{figure1}(b)). The input and detected optical powers are the same as in the previous section. The results are similar to the ones described at $\Delta/2\pi=+1.5\,\mathrm{GHz}$. The spin noise can at first well be fitted up to $n=10^{14}\,\atcm$ with a simple linear regression (blue dashed line), while a strong deviation from the linear fit is observed at higher density.

Interestingly, such a non-linear behavior is not observed for another set of larger detunings $\displaystyle{\Delta/2\pi = -7\,\mathrm{GHz}}$ and $\displaystyle{\Delta/2\pi = +8\,\mathrm{GHz}}$, shown in Fig.\,\ref{figure4}\,(b) in green squares and orange dots respectively. These values correspond to a probe beam being respectively far red and blue-detuned from all hyperfine transitions of both rubidium 85 and 87. In this case, a linear fit to the spin noise variance for densities lower or equal to $\displaystyle{n=10^{14}\,\atcm}$ (shown in green dash-dotted line and red dashed line respectively) correctly predict the spin noise variance even at higher density. This is emphasized in the inset of Fig.\,\ref{figure4} (b), where residuals from the linear fit for the $\displaystyle{\Delta/2\pi = -7\,\mathrm{GHz}}$ data are all zero within measurement errors. Note that the measured slope is much weaker than in Fig.\,\ref{figure3} (a) and \ref{figure4}\,(a) due to the larger detuning to all atomic hyperfine transitions. This linearity over more than two orders of magnitude for a far-detuned probe beam confirms the role of a light-induced excitation in the anomalous scaling shown in the previous sections.

\begin{figure}[t]
    \centering
    \includegraphics[width=\columnwidth]{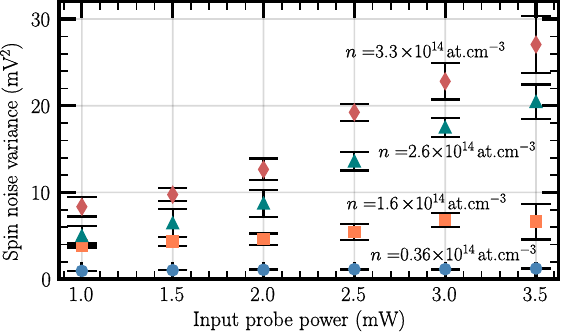}
    \caption{\textbf{Evolution of the measured spin noise on the input probe power.} Detected spin noise variance as a function of input probe power for different vapor densities. The detected power is fixed to $160\,\unit{\micro\watt}$ for all points, and the detuning is $\Delta/2\pi=+1.5\,\mathrm{GHz}$.}
    \label{figure5}
\end{figure}

\subsection{Impact of the optical probe power}
We now investigate the role of the input probe power, which we expect to enhance light-driven dipole-dipole interactions.  Let us remind here that we keep the detected power constant throughout the measurements in a same set of data, by rotating a tunable neutral density filter placed after the cell. As a result, the integrated noise variance is not expected to depend on the input optical probe power.

This assumption has been tested by measuring the spin noise variance for a fixed detuning $\displaystyle{\Delta/2\pi=+1.5\,\mathrm{GHz}}$ as a function of an input probe power ranging from $1\,\mathrm{mW}$  to $3.5\,\mathrm{mW}$, at different atomic densities. As compared to the previous sections, the detected power $P_d$ has to be lowered to $160\,\unit{\micro\watt}$  due to the weak transmitted power for $P_p=1\,\mathrm{mW}$ at high density. Results are shown in Fig.\,\ref{figure5}. For the lowest density $\displaystyle{n=3.6\times10^{13}\,\atcm}$ (blue dots), the noise variance indeed remains constant whatever the input probe power, as expected. However, for a higher density value of $\displaystyle{1.6\times10^{14}\,\atcm}$ (orange squares), a net increase in the detected signal is visible. This is even more pronounced as the density reaches $\displaystyle{2.6\times10^{14}\,\atcm}$ (green triangles) and $\displaystyle{3.3\times10^{14}\,\atcm}$ (red diamonds). This shows that the quadratic spin noise contribution present at high densities increases with the probe laser power, even though the amount of light reaching the detector is kept constant. This observation is in strong disagreement with the expected constant behavior for independent atoms, and confirms that the quadratic spin noise is indeed due to light-induced interactions at high density.

\subsection{Interpretation based on a simple model of spin correlations}


We now discuss our experimental results. First, we note that since we probe spin noise indirectly via Faraday rotation fluctuations, our observations could in principle stem from the complex optical response of the vapor at high density rather than the spin dynamics itself. Although the vapor density is a convenient experimental parameter to investigate atom-atom interactions in hot vapors, the optical depth experienced by the detuned probe at the highest densities reaches one. Multiple light scattering in this regime is no longer negligible but remains relatively weak. The following discussion therefore neglects the effect of multiple scattering of light on the detected signal and the atomic dynamics. On the other hand, spontaneously emitted photons escaping the cell are not phase related to the transmitted beam, so that their contribution to the balanced homodyne signal can be safely discarded.

Therefore, we attribute the non-linear spin noise signal to correlations between atoms at high density. Indeed, the Faraday rotation angle at the cell output is
\begin{equation}
    \varphi=g \sum_{i=1}^N \braket{s_z^{(i)}},
\end{equation}
where the sum runs over the $N$ probed atoms, and $\braket{s_z^{(i)}}$ is the quantum average $z$-component of the single-atom electronic spin operator. Here $g=\dfrac{d_0^2k}{6\varepsilon_0\hbar \Delta\Sigma}$, with $\Sigma$ the probe beam area, is an optical coupling factor originating from the atomic polarizability (see appendix \ref{app_SN_calculations}). In this section, we overlook the nuclear spin degree of freedom and the resulting hyperfine structure to simplify the discussion.
In a non-magnetized ensemble of atoms experiencing independent stochastic trajectories, spin noises incoherently add up, so that the variance of the Faraday rotation angle is
\begin{equation}\label{eq_indep_spins}
    \braket{\varphi^2}=g^2 \sum_{i=1}^N \braket{(s_z^{(i)})^2} \propto n \braket{s_z^2},
\end{equation}
where $\braket{s_z^2}$ denotes the single-particle variance of all atoms in an identical thermal state. For a small perturbation induced by the probe beam, as in our experiment, one can assume the ensemble close to thermal equilibrium. The variance $\braket{s_z^2}$ is then solely determined by the ground-state properties and the vapor temperature \cite{mouloudakis_effects_2022}: $\braket{s_z^2}=\mathrm{Tr}\left(\rho_{th}^{(1)}s_z^2\right)$ where $\rho_{th}^{(1)}$ is the single-body thermal state density matrix. Since $\braket{s_z^2}$ is independent of the density, according to eq.(\ref{eq_indep_spins}) the resulting detected spin noise is linear with $n$, in agreement with our experimental results at low vapor densities.

At higher density, if atomic interactions are strong enough to induce correlations between individual spin dynamics, eq.\,(\ref{eq_indep_spins}) has to be modified to take into account pair-wise cross-correlations
\begin{equation}\label{eq_corr_spins}
    \braket{\varphi^2}=g^2\left[ \sum_{i=1}^N \braket{(s_z^{(i)})^2} + \sum_i ^N\sum_{j\neq i}^{N}\frac{\braket{s_z^{(i)}s_z^{(j)} + s_z^{(j)}s_z^{(i)}} }{2} \right].
\end{equation}
Although the amplitude of the two-body correlations certainly depends on $n$, owing to the Cauchy-Schwartz inequality it admits the constant single-particle variance as an upper bound: $\displaystyle{\braket{s_z^{(i)}s_z^{(j)} + s_z^{(j)}s_z^{(i)}} \leq \braket{2(s_z^{(i)})^2}}$. The second term in eq.\,(\ref{eq_corr_spins}) is thus expected to maximally grow as the squared atomic density $n^2$. At sufficiently high densities, the signal is thus expected to follow
\begin{equation}\label{eq_theory_high_density}
    \braket{V^2} \propto g^2\left( \beta n+ \gamma n^2\right),
\end{equation}
where $\beta$ describes the \textit{uncorrelated} part of the noise and is related to the equilibrium state of the vapor, and $\gamma$ parametrizes the \textit{correlated} fraction of the noise. This simple model matches our experimental observation of an additional non-linear density contribution to the spin noise variance. Moreover, our results suggest that the cross-correlation coefficient $\gamma$ depends on the vapor optical excitation via the light Rabi frequency $\Omega$ and detuning $\Delta$ to nearby atomic transitions: $\gamma=f(\Omega, \Delta)$.

Nevertheless, the fact that the non-linear spin noise scaling is only observed for densities higher than $\displaystyle{n_c\simeq 10^{14}\,\atcm}$ is not inherently captured by the simple model described above. As noted in section \ref{section1_exp}, the spin noise variance cannot be fitted by a single equation such as eq.\,(\ref{eq_theory_high_density}) across the whole density range.  Instead, we resort to the discontinuous model described above, where we first use a linear regression with a slope $\kappa_{exp}$ before adding a quadratic contribution $\propto (n-n_c)^2$ at higher densities. Given our current experimental precision, whether this density threshold effect results from a genuine non-analytical behavior rather than a sharp density dependence of the correlation coefficient $\gamma$ cannot be definitively settled. 

In both cases, our experimental results emphasize that the high-density spin noise is intrinsically many-body, with \textit{positive} correlations between atoms in this case. Reproducing these results requires a many-body noise model beyond the widely-used mean-field Bloch equations \cite{mouloudakis_effects_2022} and the two-body model introduced in \cite{delpy_anomalous_2025}. Moreover, deeper analysis of the latter revealed an important discrepancy between the amplitude of the measured and predicted low-frequency broadband noise (described in section \ref{section1_exp}), which also suggests the limit of the two-atom picture. Going beyond this framework to describe correlations requires to explicitly model the dynamics of a many-body ensemble interacting with light, a theoretical challenge we let for future work.

In order to bring additional experimental evidence to this discussion, the next section introduces a protocol aiming at tuning the strength of the dipole-dipole interactions. In contrast with the previous sections, we thus investigate changes in the non-linear spin noise contribution at a constant atomic density, while the excited-state population is quenched by a second laser.

\section{Suppression of correlations using an auxiliary resonant beam}\label{section3}

\subsection{Excitation scheme and protocol}

Suppression of resonant dipole-dipole interactions is achieved by depopulating the 5$^2$P$_{3/2}$ excited state through the 776 nm 5$^2$P$_{3/2} \rightarrow$ 5$^2$D$_{5/2}$ transition. The relevant energy levels scheme is shown in Fig.\,\ref{figure1}(b). We use a second Toptica ECDL for the auxiliary beam, distinct from the probe beam. This auxiliary laser is tuned to the approximate center of gravity of the fine-structure transition in order to depopulate all the hyperfine levels of 5$^2$P$_{3/2}$. A QWP removes any ellipticity in the beam and a HWP sets the linear polarization orthogonal to that of the probe. The auxiliary beam is sent through the cell at a small angle relative to the pump beam. Inside the cell, the auxiliary beam is focused to a diameter $\phi_a \simeq$ 400 $\mu$m, ensuring a complete spatial overlap with the probed area. The transmitted power of the auxiliary beam is detected by a Thorlabs PM100D power meter.

\begin{figure}
    \centering
    \includegraphics[width=\columnwidth]{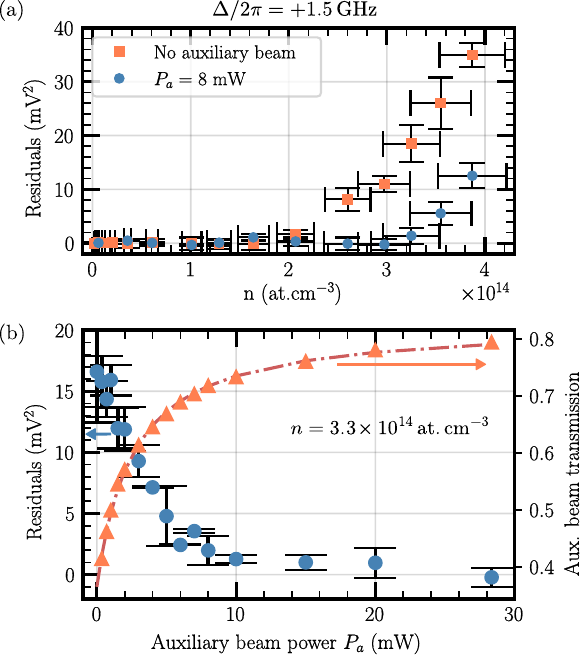}
    \caption{\textbf{Impact of the auxiliary beam on the detected spin noise variance.} (a) Density dependence of the residuals from the low-density linear trend with (blue dots) and without (orange squares) auxiliary beam. Parameters are $P_p=2\,\mathrm{mW}$, $P_d=0.4\,\mathrm{mW}$ and $\Delta/2\pi=+1.5\,\mathrm{GHz}$. (b) Blue dots: evolution of the residuals with auxiliary beam power at a fixed density of $3.3\times10^{14}\,\atcm$. Orange triangles: transmission of the auxiliary beam, fitted with a saturated optical depth function (red dash-dotted line).}
    \label{figure6}
\end{figure}

\subsection{Evolution of the spin noise variance and lineshape with the auxiliary beam power}

\paragraph{Suppression of the quadratic spin noise component.} The impact of the excited-state depopulation due to the auxiliary beam is shown in Fig.\,\ref{figure6}. We fix all parameters as in Fig.\,\ref{figure3}: $P_p=2\,\mathrm{mW}$, $P_d=0.4\,\mathrm{mW}$ and $\Delta/2\pi=+1.5\,\mathrm{GHz}$.
The residuals of the spin noise variance from the low-density linear fit is shown as a function of the vapor density for two cases: without the auxiliary beam (orange squares, same data as in Fig.\,\ref{figure3}), and with an $8\,\mathrm{mW}$ auxiliary beam (blue dots). Interestingly, in the latter case the residuals are zero up to a density of $3\times 10^{14}\,\atcm$, twice as much as compared with the "probe laser only" configuration. The spin noise linearity is therefore partially retrieved: at the highest density, the non-linear contribution is reduced by a factor of three by the auxiliary beam.

More insight on this noise reduction can be obtained by analyzing the impact of the auxiliary beam power, which governs the fraction of the initial excited-state population which is pumped away, at a fixed density of $3.3\times10^{14}\,\atcm$ ($T=175^{\circ}\mathrm{C}$). The results are shown in Fig.\,\ref{figure6}\,(b) (blue dots), along with the measured transmission of the auxiliary beam (orange triangles), which reflects changes in the population of the first excited state. When increasing the auxiliary beam power, the spin noise residuals drop quickly, and the transmission of the auxiliary beam is simultaneously enhanced. This suggests that as the 776 nm beam depopulates the $5^2\,\mathrm{P}_{3/2}$ state, the non-linear scaling of the spin noise is suppressed. We fit the transmission data with a saturated transmission function $\displaystyle{f(P_a) = L\exp\left({-\frac{\alpha_0}{1+P_a/P_s}}\right)}$, where $\alpha_0$ is an unsaturated optical depth on the 5$^2$P$_{3/2} \rightarrow$ 5$^2$D$_{5/2}$ transition, $P_s$ a saturation power and $L$ losses due to the cell windows. We find a typical saturation power of 1.7 mW. Consequently, for beam powers $P_a>10\,\mathrm{mW}$, the vapor is practically transparent to the 776 nm beam and the spin noise residuals tend to zero. This again confirms that the linear scaling of spin noise with vapor density is restored when the excited-state population is quenched. On the contrary, we verified that at low density, i.e. in the non-interacting regime, the auxiliary beam power has no impact on the measured noise variance.

\begin{figure}
    \centering
    \includegraphics[width=\columnwidth]{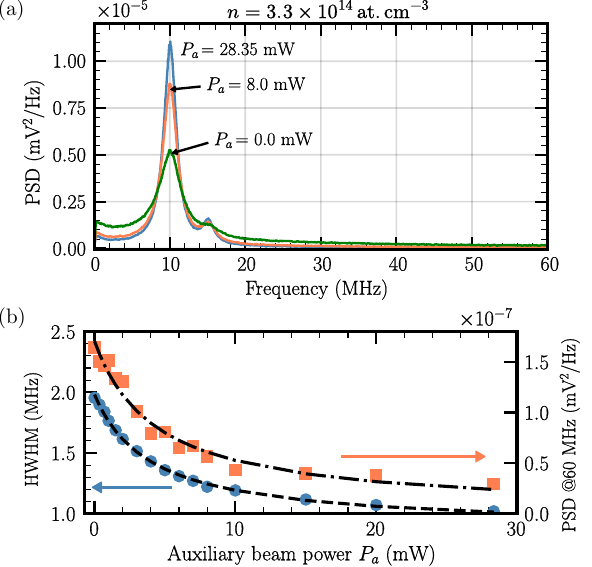}
    \caption{\textbf{Impact of the auxiliary beam power on the lineshapes.} (a) Modification of the spin noise lineshapes with increasing auxiliary beam power depopulating the excited state. (b) Reduction of $^{85}$Rb spin linewidth (blue dots) and broadband low-frequency noise measured at 60 MHz (orange squares). Dashed and dash-dotted line: fit by a saturation function.}\label{figure7}
\end{figure}

\paragraph{Changes in lineshape.}
Finally, the spectral analysis of the detected signals allows us to correlate such results with a frequency redistribution of the spin noise. Figure \ref{figure7}\,(a) indeed shows a clear reshaping of the spin noise spectra as the auxiliary beam power is increased from 0 to 28 mW. Specifically, one observes (i)\textit{ a narrowing} of the spin noise peaks, and (ii)\textit{ a reduction} of the broadband low-frequency noise, that is, a suppression of the hallmarks of the resonant dipole-dipole interaction \cite{delpy_anomalous_2025}.

This is confirmed by Fig.\,\ref{figure7} (b), which shows a reduction of the HWHM of the $^{85}\mathrm{Rb}$ peak from almost 2 MHz to 1 MHz (see blue dots). Such linewidths are obtained by locally fitting the SN peak centered on 10~MHz. This decay is well captured by a fitting function $f(P_a)= \gamma_0 + \dfrac{\delta \gamma}{1+P_a/P_s}$, where $\gamma_0$ is a supposedly unperturbed linewidth, $\delta\gamma$ is the interaction-induced broadening and $P_s$ a saturation power of the excited-state transition in our experimental conditions. Such fitting function (in black dashed line) yields a saturation power of $4.0\pm0.2\,\mathrm{mW}$, larger but of the same order of magnitude as the one extracted from the transmission of the auxiliary beam. The unperturbed linewidth $\gamma_0$ is around $0.9\,\mathrm{MHz}$, larger than the transit-limited $0.25\,\mathrm{MHz}$ linewidth, suggesting that a complete depletion of the excited states is not achieved due to the inhomogeneous Doppler broadening.

Additionally, a reduction of the broadband noise component centered on low frequencies is demonstrated by measuring the noise PSD level at a fixed frequency of $\displaystyle{60\,\mathrm{MHz}}$, a region of the spectrum in which the SN peaks of both isotopes do not contribute. This noise level, shown in orange squares in Fig.\,\ref{figure7} (b), steadily decreases as the auxiliary beam power is increased. A similar fitting function as used for the linewidth yields a saturation power $P_s=4.3\pm 0.8\,\mathrm{mW}$ in perfect agreement with the previous estimation. Eventually, the fact that both the spin noise non-linearity and lineshape distortions are suppressed by the auxiliary beam supports our claim that both features originate from dipole-dipole interactions in the vapor.

\section{Conclusion}

We have reported a previously unobserved non-linear spin noise in a dense rubidium vapor, contrasting with the usual proportionality between spin variance and atomic density in dilute media. We have shown that a quadratic component in the spin variance appears for densities higher than $10^{14}\,\atcm$, together with strong distortions of the spin noise spectrum attributed to light-induced dipole-dipole interaction in an earlier work. The impact of the optical excitation on the quadratic spin noise was confirmed by a strong dependence on the probe light power and detuning to atomic transitions between the 5$^2$S$_{1/2}$ and 5$^2$P$_{3/2}$ levels.

Based on our observations, we suggest that this anomalous scaling is due to positive atom-atom correlations arising due to resonant dipole-dipole interactions. Such interpretation is supported by a suppression of the non-linear contribution to the total spin noise when inhibiting this interaction mechanism. This was achieved by tuning a second laser beam on the 5$^2$P$_{3/2} \rightarrow$ 5$^2$D$_{5/2}$ excited state transition, thereby depopulating the first excited level. We experimentally demonstrated that a linear scaling of spin noise can be restored when saturating the absorption of the auxiliary beam, indicating an efficient reduction of atomic correlations.

These observations highlight the many-body nature of spin noise in physical systems where interactions allow inter-particle correlations to build up.
Moreover, our experimental observations suggest that the non-linear spin noise variance can be detected only beyond a critical threshold density, a feature often associated to phase transitions. Interestingly, Gorshkov \textit{et al.} theoretically proved that spin-spin interactions could result in phase transition-like changes in spin noise dynamics \cite{gorshkov_phase_2024}, with an additional low-frequency noise component as a corresponding order parameter. Such predictions remarkably reminds of the experimental results reported in \cite{delpy_anomalous_2025} and in this article. Investigating any formal analogy between the two frameworks is thus an exciting perspective. In our case, which involves light-induced processes rather than direct spin interactions, the role of multiple photon scattering at unit optical depths should also be clarified.

Another natural and interesting follow-up prospect is the clarification of the quantum or classical statistics of such correlations. This is usually done by comparing spin variance with relevant lower-bound inequalities \cite{geza_PRA}, with the additional difficulty of having here positive correlations rather than the negative ones usually associated to spin squeezing \cite{kong_measurement-induced_2020}. Whether spin fluctuations are gaussian or not in the correlated regime is an important related question, and could in principle be investigated using higher-order SNS \cite{sinitsyn_theory_2016, li_higher-order_2013, Li_Sinitsyn_2016}. Solving these questions could improve our knowledge of collective effects in dense atomic vapors and spin dynamics in open, many-body-correlated ensembles.

\acknowledgments
The authors thank Antoine Browaeys and Thomas F. Cutler for their help with the vapor cell, Ehouarn Le Roy and Jiong Deng for their experimental support and from S\'ebastien Rousselot for technical assistance. This work benefited from useful discussions with Kostas Mouloudakis and Jan Ko\l{}ody\'nski. The authors warmly thank Iannis Kominis for his insight and suggestions regarding the protocol described in section \ref{section3}. The authors acknowledge funding by the Labex PALM.

\appendix

\section{Background subtraction and integration procedure}\label{app_background}

\subsection{Background subtraction.}
We describe here the procedure for extracting and analyzing the spin-noise signal obtained from the digital oscilloscope. We restrict ourselves to a frequency range of 0.2 MHz to 250 MHz. Below 0.2 MHz, flicker $\left( 1/f \right)$ noise dominates the spectrum. The upper frequency is chosen appropriately based on the 350 MHz bandwidth of the balanced photodetector.

In order to properly evaluate the integral of the spin-noise PSD, accurate subtraction of the background is imperative. In our experiment, the background is composed of two contributions: (a) the dark background of the balanced photodetector, amplification chain and oscilloscope, and (b) the laser shot-noise. We systematically subtract these two backgrounds using the following procedure:
\begin{enumerate}
    \item We first acquire the dark electronic background noise $S_{elec} \left( \nu\right)$, with the laser turned off. This includes the technical background noise of the whole detection chain.
    \item We then acquire the background $S_{bg}$ with the laser on but no vapor cell. This contains both the laser shot-noise $S_{laser} \left( \nu \right)$, and the dark electronic background: $S_{bg} \left( \nu \right)~=~S_{elec}\left( \nu \right)~+~S_{laser} \left( \nu \right)$. The laser shot-noise can be thus accessed by performing the subtraction $\displaystyle{S_{laser} = S_{bg} - S_{elec}}$. The orange curve labeled "Background noise" in Fig.\,\ref{fig:app2_bg_sub} shows a typical background signal extending across the full-range of measured frequencies.
\end{enumerate}

Subsequently, the total measured signal, $S_{meas}$, is a combination of the electronic background, the laser shot-noise and the actual atomic response itself. Two such curves measured at 120$^{\circ}$C and 177$^{\circ}$C are shown in Fig.\,\ref{fig:app2_bg_sub}. The inset figure shows a zoomed-in view up to 30~MHz, highlighting the main spin-noise peaks. The spin noise signal $S_{SN}$ is retrieved from the measured signal by subtracting the total background, i.e., $\displaystyle{S_{SN} = S_{meas} - S_{bg}}$. This final form of the PSD is shown in other parts of the text [Fig.\,\ref{figure2}, Fig.\,\ref{figure7}\,(a)].

\subsection{PSD integration procedure.}
\begin{figure}
    \centering
    \includegraphics[width=\columnwidth]{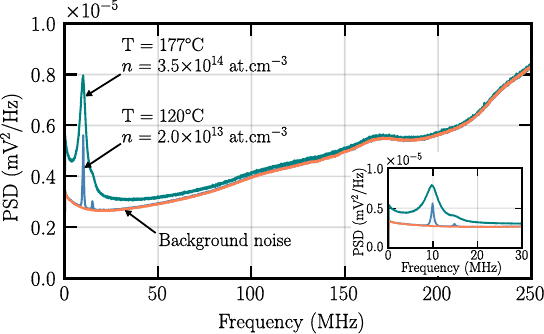}
    \caption{\textbf{Examples of raw PSD along with total background noise.} Integration bounds for spin noise variance calculations are determined by the frequencies around which the total PSD merges with the background.}
    \label{fig:app2_bg_sub}
\end{figure}

In order to obtain the spin-noise variance, $S_{SN}$ is then integrated by manually choosing an appropriate frequency domain. The total spin-noise variance is composed of contributions from the resonance peaks of $^{85}$Rb and $^{87}$Rb as well as the broadband noise component appearing at higher densities. Hence, the integration domain should appropriately account for all these contributions. At low densities, it is not necessary to integrate over all the measured frequency range, as the only contribution to the signal is from the two narrow SN peaks.
This can be seen in Fig.\,\ref{fig:app2_bg_sub} where, apart from the two sharp resonance features, the signal for $n=2.0\times 10^{14}\,\atcm$ (T=120$^{\circ}$C) is flat and overlaps with the background $S_{bg}$. Hence, the integration domain can be chosen in the near vicinity of these peaks, typically between 7 and 20 MHz.

However, for higher densities, the broadening of the peaks and the broadband noise component implies the need for a larger domain of integration (see green curve in Fig.\,\ref{fig:app2_bg_sub}). On the lower end, the minimal bound is always limited by the 0.2 MHz cut-off frequency from the high-pass filtering. The upper bound is then chosen where $S_{SN}$ significantly merges with $S_{bg}$ for each spectrum, within the maximum limit of 250 MHz, smaller than the finite balanced detection bandwidth.

\begin{figure}
    \centering
    \includegraphics[width=\columnwidth]{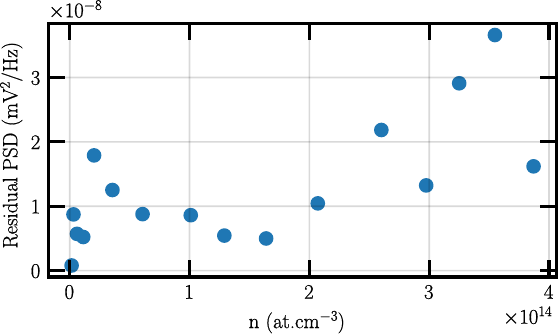}
    \caption{\textbf{Examples of residual PSD levels after background subtraction,} from which uncertainties in the estimated spin noise variance are derived.}
    \label{fig:app2_PSD_residuals}
\end{figure}

To further quantify any potential residual background or spin noise power left out during integration, we average the PSD $S_{SN}$ in a 10 MHz range around the higher cut-off frequency. An example of the obtained residual PSD are shown in Fig.\,\ref{fig:app2_PSD_residuals} for $\Delta/2\pi = +1.5\,\mathrm{GHz}$ and a detection power of $400\,\unit{\micro\watt}$. A roughly uniform scatter at all densities indicates that no significant power is being inadvertently lost when choosing the integration domain. The remaining PSD level can then be imputed to an imperfect background subtraction, and is used to derive the errors bars in the spin-noise variance plots in the main text.

\section{Linearity of the photodetection}\label{app_linearity}
We experimentally show here that the observed non-linearities and lineshape changes at higher densities are not an artifact of the photodetector and that all the measurements are performed within the linear regime of the detector. 

\begin{figure}
    \centering
    \includegraphics[width=\columnwidth]{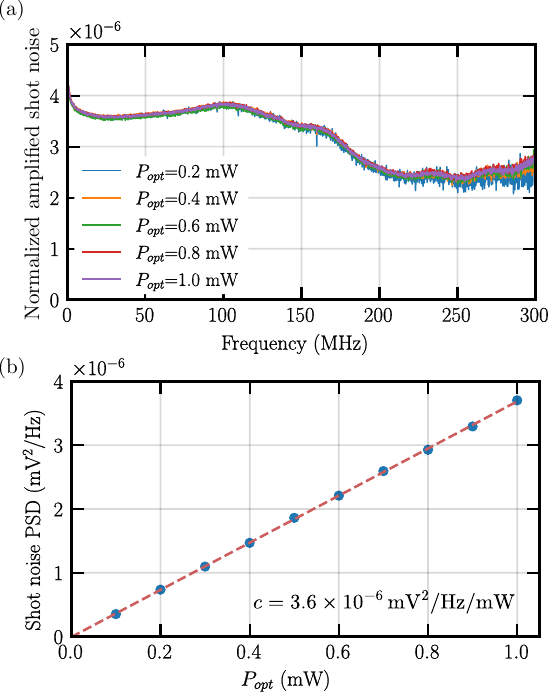}
    \caption{\textbf{Characterization of the photon shot noise power spectrum.} (a) PSD of the detected shot noise for different optical powers, after normalization by this power. (b) Linearity of the shot noise PSD evaluated at a fixed 10 MHz frequency with input optical power.}
    \label{fig:app1_lin_PSD}
\end{figure}

We measure the laser shot-noise signal at different optical powers, while balancing the detection for each measurement, by getting rid of the vapor cell on the beam path. Figure \ref{fig:app1_lin_PSD}\,(a) shows the detected shot noise PSD $S_{laser}(\nu)$ after subtraction of the electronic background (see Appendix \ref{app_background}), at five different optical powers $P_{opt}$ ranging from 0.2 mW to 1.0 mW. Our usual detection power of 0.4 mW lies well within this range. Since we expect a shot noise power $S_{laser}\propto P_{opt}$, each curve is normalized by dividing the corresponding optical power. It is clear that the scaled PSDs at all measured optical powers fall on top of each other, indicating a good linearity of our photodection scheme over a 300 MHz frequency range. Note that since photon shot noise is a white noise, Fig.\,\ref{fig:app1_lin_PSD}\,(a) also gives away the frequency response function of the detection and amplification chain, which is used to normalize the computed spin noise spectra.

Furthermore, Fig.\,\ref{fig:app1_lin_PSD}\,(b) shows the shot-noise level measured at 10 MHz as a function of the optical power. This frequency is chosen to show the linearity of the PSD near the resonance peaks within the area of interest. A linear fit gives a slope of $s_{exp} = 3.6 \times 10^{-6}$ mV$^2$/Hz/mW. The theoretically expected slope calculated using the datasheet of Thorlabs PDB435A at a transimpedance gain of 10$^4$ V/A and a responsivity of 0.47 A/W at 780~nm is precisely $3.6\times 10^{-6}\, \mathrm{mV}^2$/Hz/mW, in excellent agreement with the experimental data.

\section{Calculations of absolute Faraday rotation noise variance} \label{app_SN_calculations}

We consider an ensemble of atoms having a simple electronic spin degree of freedom in the ground state. Let us assume that this ensemble has a net magnetization in the $z$ direction, denoted $\braket{S_z} = \dfrac{1}{N} \sum_{k}\braket{s_z^{(k)}}$, where $N$ is the number of probed atoms and $s_z^{(k)}$ the single-particle spin operator of atom $k$. A linearly polarized probe beam propagating along $z$ in the vapor while being tuned near an atomic resonance will experience a polarization rotation by an angle $\varphi$ due to the Faraday effect. Denoting $d_0$ the dipole matrix element of the nearby transition, $\Delta$ the detuning, $k$ the light wave vector and $L$ the cell length, it is given by \cite{vasilakis_precision_2011}
\begin{equation}\label{eq_FR_angle}
    \varphi = \dfrac{d_0^2knL}{6\varepsilon_0\hbar \Delta} \braket{S_z},
\end{equation}
in the relevant limit $\Delta \gg \Gamma_0, \Gamma_D$ where $\Gamma_0$ denotes the excited state population decay rate and $\Gamma_D$ the Doppler inhomogeneous broadening. Note than in section \ref{section1_exp} for instance, we typically have a detuning $\Delta \simeq 5\times \Gamma_D$ to the closest hyperfine transitions of $^{85}\mathrm{Rb}$, hence we are not strictly in the limit $\Delta \gg \Gamma_D$. We still chose to overlook the Doppler broadening and use eq.\,\ref{eq_FR_angle} here, since these calculations only aim at providing an estimation of the spin noise level to validate our experimental protocol.

In our case, the cell contains two isotopes (labeled by the superscript $(i)$), showing spin-orbit and hyperfine coupling due to their nuclear magnetic moment $I^{(i)}$. One must now replace the electronic spin operator by the set of hyperfine spin operator $F_z$, and take into account the detuning $\Delta_F$ to each transition between a ground-state hyperfine $F$ level and the excited state. Here we neglect the excited-state hyperfine splitting, which is unresolved due to Doppler broadening in warm rubidium vapors. The Faraday rotation angle then writes
\begin{equation}
    \varphi = \dfrac{d_0^2knL}{6\varepsilon_0\hbar} \sum_i\sum_F \dfrac{A_i}{2I^{(i)}+1} \dfrac{\braket{F_z^{(i)}}}{\Delta_F}.
\end{equation}
where $n$ is the total atom density and $A_i$ is the abundance of isotope $i$.
We use the definition of $d_0$ given by $\displaystyle{\Gamma_0 = \dfrac{d_0^2\omega_0^3}{3\hbar\varepsilon_0c^3} \dfrac{2J+1}{2J'+1}}$, with $c$ the speed of light in vacuum, $\omega_0$ the transition frequency between ground and excited states with respective electronic angular momentum $J$ and $J'$. For the $\mathrm{D}_2$ transition of rubidium, $d_0 = 3.58\times 10^{-29}\,\mathrm{C.m}$ \cite{steck_rubidium_2001}.

In SNS experiments, the vapor is unpolarized, so that $\braket{F_z^{(i)}}=0$ for every $i$ and $F$. We measure the variance of the Faraday angle $\braket{\varphi^2}$
\begin{equation}
    \braket{\varphi^2} = \left[\dfrac{d_0^2knL}{6\varepsilon_0\hbar}\right]^2 \sum_i\sum_F \dfrac{A_i^2}{[2I^{(i)}+1]^2} \dfrac{\braket{(F_z^{(i)})^2}}{\Delta_F^2}. \label{eq_FRvariance}
\end{equation}
The variance of the $z$-component of all hyperfine spin operators writes \cite{mouloudakis_effects_2022, shah_high_2010}
\begin{equation}
    \braket{(F_z^{(i)})^2} = \dfrac{1}{N_i}\dfrac{F(F+1)(2F+1)}{6(2I^{(i)}+1)}=\dfrac{\braket{f_z^2}}{N_i},
\end{equation}
where $f_z$ is a single-atom hyperfine spin operator. We used here the assumption of a mixed state $\rho_N=\rho_{th}^{\otimes N}$, with $\rho_{th}$ the single-atom thermal state, to describe a non-interacting ensemble of atoms at high temperature. Note that we already neglected correlations between different hyperfine levels in eq.(\ref{eq_FRvariance}), in agreement with the hypothesis of an uncorrelated ensemble.

The total number of probed atoms being $N=nL\Sigma$, where $\Sigma$ denotes the effective probe beam area, the total Faraday rotation noise variance finally writes
\begin{equation}
    \braket{\varphi^2} = \alpha^2 \dfrac{nL}{\Sigma} \sum_i\sum_F \dfrac{A_i}{[2I^{(i)}+1]^2} \dfrac{\braket{(f_z^{(i)})^2}}{\Delta_F^2},
\end{equation}
where $\alpha= \dfrac{d_0^2k}{6\varepsilon_0\hbar}$ is an optical coupling factor reminiscent of the atomic polarizability. 

Let us emphasize here that the proportionality of $\braket{\varphi^2}$ with the atom density $n$ directly originate from the approximation of uncorrelated atoms. It is this approximation which we believe breaks down at higher density and could explain the results reported in this article.

Finally, taking into account the optically detected power $P_d$, the balanced detection sensitivity $\eta$, transimpedance $R_t$ and an additional amplification gain $g_{dB} = 23.8\,\mathrm{dB}$, the experimentally measured voltage variance is
\begin{equation}
    \braket{V^2}= \left[ 2\eta R_t P_d\times 10^{\frac{g_{dB}}{20}}\right]^2 \braket{\varphi^2}
\end{equation}
These equations are used with the relevant beam size parameters and electronic specifications to calculate the theoretical voltage variance in the experiment presented in section \ref{section1_exp}.

\bibliographystyle{unsrt}

\end{document}